\documentclass{article}

\usepackage{arxiv}

\usepackage[utf8]{inputenc} 
\usepackage[T1]{fontenc}    
\usepackage{url}            
\usepackage{booktabs}       
\usepackage{amsfonts}       
\usepackage{nicefrac}       
\usepackage{microtype}      
\usepackage{lipsum}		
\usepackage{graphicx}
\usepackage[square,sort,comma,numbers]{natbib}
\usepackage{doi}

\usepackage{hyperref}   

\title{Interpretation of immunofluorescence slides by deep learning techniques:  anti-nuclear antibodies case study}

\author{ \hspace{1mm}Oumar Khlelfa\\
	Military Academy\\
	Fondouk Jedid, Nabeul 8012, Tunisia \\
	\texttt{oumarkh19971@gmail.com} \\
	\And
	\hspace{1mm} Aymen Yahyaoui  \\
	Military Academy\\
	Fondouk Jedid, Nabeul 8012, Tunisia\\ 
 Science and Technology for Defense Lab (STD), Ministry of National Defense,
    Tunisia \\
	\texttt{aymen.yahyaoui@ept.rnu.tn} \\
	  \AND
	Mouna Ben Azaiz \\
	Immunology Department, The Principal Military      Hospital of Instruction of Tunis\\
	Tunisia \\
	  \And
	Anwer Ncibi \\
	Military Academy \\
	Fondouk Jedid, Nabeul 8012, Tunisia\\
	\And
	Ezzedine Gazouani \\
	Immunology Department, The Principal Military      Hospital of Instruction of Tunis \\
	Tunisia \\
        \And
        Adel Ammar\\
        Robotics and Internet-of-Things Laboratory, Prince Sultan University, Riyadh, Saudi Arabia\\
        \texttt{aammar@psu.edu.sa }
        \And
        Wadii Boulila\\
        Robotics and Internet-of-Things Laboratory, Prince Sultan University, Riyadh, Saudi Arabia\\
        RIADI Laboratory, National School of Computer Sciences, University of Manouba, Manouba, Tunisia\\
        \texttt{wboulila@psu.edu.sa}
        }

\renewcommand{\shorttitle}{\textit{arXiv} Template}

\begin{document}
\maketitle

\begin{abstract}
	Nowadays, diseases are increasing in numbers and severity by the hour. Immunity diseases, affecting  8\% of the world population in 2017 according to the World Health Organization (WHO), is a field in medicine worth attention due to the high rate of disease occurrence classified under this category. This work presents an up-to-date review of state-of-the-art immune diseases healthcare solutions. We focus on tackling the issue with modern solutions such as Deep Learning to detect anomalies in the early stages hence providing health practitioners with efficient tools. We rely on advanced deep learning techniques such as Convolutional Neural Networks (CNN) to fulfill our objective of providing an efficient tool while providing a proficient analysis of this solution. The proposed solution was tested and evaluated by the immunology department in the Principal Military Hospital of Instruction of Tunis, which considered it a very helpful tool.
\end{abstract}


\section{Introduction}
In the current era, the significant spread of chronic diseases is much higher than ever before, especially those related to immune problems~\cite{ref_article89}. On this note, the fact that the rate of affection by immune diseases is increasing extensively underlines the need to provide a solution for the doctors to help save time and provide them with precision while treating the immunofluorescence slides.
\quad In fact, the analysis of immunofluorescence~\cite{ref_article90} results based solely on the knowledge of a single doctor is a method that lacks precision, and also could be a significant source of potential misdiagnosis due to the massive amount of details present in the Indirect immunofluorescence (IIF) images~\cite{ref_article91}. This creates dangerous risks for immunologists working under enormous responsibility and likewise for the patient, who is the most significant beneficiary.
Therefore, the presented solution is very effective in this matter, as it helps to save time, leading to faster results; thus, more patients can be treated by a single expert doctor. On the other hand, the certainty factor also plays a major part in this study, as the error rate is very minimal thanks to this model, which made it possible to reach a 94.48\% accuracy for diagnosing positive and negative IIF images. This will lead to another level of performance that will help newly graduated doctors to go through their medical career with more confidence based on the automated results.
\quad This work is carried out in collaboration with the immunology department of the Principal Military Hospital of Instruction of Tunis. The objective is to offer a tool in the form of a computer application, where the greatest benefits are to effectively assist the doctors of the immunology department in diagnosing autoimmune diseases through the analysis of IIF images.
 This should make interpreting IIF images much easier in terms of time and level of certainty.
 
 The main contributions of this work are as follows: 
 \begin{itemize}
     \item It presents an overview of the different state-of-the-art techniques, methods, and tools related to AI and the interpretation of immunofluorescence sides.
     
     \item It treats the multiple cases of immunofluorescence slides using a combination of a DL model and a NASNet model leading to accurate results.
     
     \item It presents the interactive tool developed for the specific needs of the medical staff at the military hospital of Tunis immunology department. This tool was developed, deployed, and tested on many patients. The medical staff confirmed it as an interesting and useful tool for IIF interpretation.
  \end{itemize}   
  
The science of immunology was revealed in the late 18th century, and from that moment on, it can be admitted that it gained its rightful recognition as a branch of knowledge.\newline
Autoimmune diseases,
 one of the most common issues in immunology, is a disease that results from a dysfunction of the immune system leading it to attack the normal constituents of the body: Normally, the immune system preserves the cells of the body. Whereas during autoimmune disease, it identifies them as foreign agents and attacks them. Statistics have shown that 5 to 8\% of the world's population is affected by this type of disease~\cite{ref_article2}.\newline
Immunofluorescence technique: 
 It is a method that consists of detecting and locating, by fluorescence emission, a protein of interest produced by an Antigen, with the help of a specific antibody related to the agent in question. Thus, it makes it possible to determine the presence or absence of a protein and its localization in the cell or tissue analyzed. 
 In immunofluorescence, two types of labeling can be carried out as shown in Figure 1:
 \begin{itemize}
     \item Direct immunofluorescence~\cite{ref_article92} where the fluorescent protein, called fluorochrome, is not coupled to a secondary antibody but to the primary antibody.
     \item Indirect immunofluorescence~\cite{ref_article90} where a primary antibody binds to the antigen. Then, a secondary antibody with a strong affinity for the primary antibody is added.
\end{itemize}
 \begin{figure}[h!]
    \centering
    \includegraphics[width=0.5\textwidth]{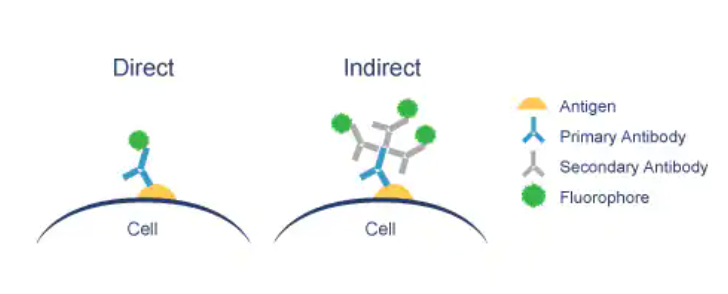}
    \caption{Direct vs Indirect Immunofluorescence~\cite{ref_dirindir}.} 
    \label{fig0}
    \end{figure}

\section{Related Work}
Indirect immunofluorescence (IIF) is considered the reference test for the detection of autoimmune diseases.
As a well-established and challenging issue in the field of medical image analysis, HEp-2 image classification has become one of the growing centers of interest in the last decade.
For this reason, three international IIF image classification competitions were held in 2012, 2014, and 2016. These competitions were very significant in this progress by facilitating the collection of a large amount of data.
As a classical image classification problem, traditional machine learning techniques have been greatly applied to HEp2 image classification.
\subsection{Machine Learning-based solutions}
In this context, several research works have been considered using various methods such as gradient features with intensity order pooling by~\cite{ref_article3} who were capable of reaching 74.39\% cell level accuracy and 85.71\% image level accuracy on the ICPR dataset. Also, multiple linear projection descriptors were considered by ~\cite{ref_article4}, who adopt the feature learning method to learn the appropriate descriptor from the image data itself. They were able to reach 66.6\% classification accuracy. Machine Learning (ML) can learn from the data, while non-learning-based techniques depend on rules that depend critically on domain knowledge. However, traditional machine learning techniques rely on predefining feature representations, which is a crucial step and necessitates complicated engineering \cite{qawqzeh2020classification}.
\subsection{Deep Learning based solutions}
Several deep learning-based approaches proposed in the literature have shown great progress in healthcare for disease diagnosis \cite{driss2020novel,al2021novel,al2021feature,rasool2022hybrid,ben2022randomly,jemmali2022equity,alam2022public}.
Deep learning was considered in two subfields of HEp2 image classification (HEp2IC) that have attracted the attention of researchers: the classification of individual HEp-2 cells and the classification of HEp-2 specimens.
\subsubsection{Cell-level HEp-2 image classification (CL-HEP2IC) methods: }
there are two methods for Deep Neural Networks (DNN) to be used; DNN as a feature extractor and DNN as a classifier. As shown in Figure 2,  the amount of research based on the last method is more important than the other. In fact, more than 30 publications adopt this method on three of the most popular datasets (ICPR 2012, I3A, and SNPHEp-2). In contrast, less than ten publications are adopting the method of using the DNN as a feature extractor.
    \begin{figure}[h!]
    \centering
    \includegraphics[width=0.5\textwidth]{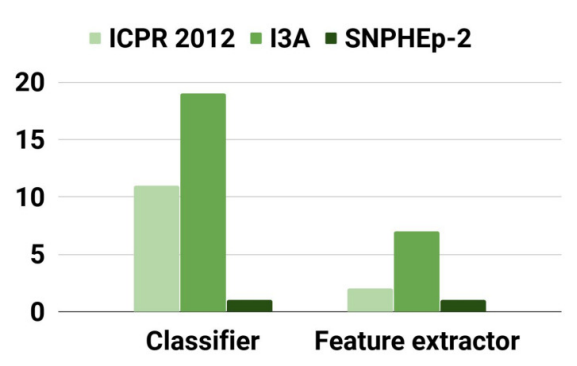}
    \caption{Number of publications based on the use of DNNs~\cite{ref_article5}.} 
    \label{fig1}
    \end{figure}

  1) CL-HEp2IC methods that use DNN as feature extractor:  \\
  Extraction is frequently used in HE2IC approaches based on handcrafted features. It allows extracting features from the input image to carry out the classification performed generally with support vector machines or k-nearest neighbors. There are two kinds of feature extraction: feature extraction from a pre-trained DNN model and feature extraction from a fine-tuned DNN model: First, in~\cite{ref_article6}, authors used a pre-trained CNN model but with an intensity-aware classification schema with two steps accompanied by an approach to pull out the features that can differentiate between the classes which allowed them to reach 77.1\% average class accuracy. One year later, authors in \cite{ref_article7} proposed a much simpler idea of replacing the SIFT with the CNN features, enabling them to achieve a 98\% classification accuracy. Second, feature extraction from a fine-tuned DNN model consists of improving a pre-trained model. One of the methods in this context involves a modern pooling strategy to overcome and remove the fixed-size constraint of the CNNs known by resizing the input images. They used a CaffeNet model to extract features and then a K-spatial pooling to support the HEp2 cell images with arbitrary sizes. This method allowed them to reach an accuracy of 98.41\%~\cite{ref_article5}. Unlike all previous works, authors of~\cite{ref_article10} used the convolutional auto-encoder (CAE) as a feature extractor. In fact, it is not only CAE, but they train two different CAEs, one for normal images with RGB colors and another one for the gradient images, which are images that have been subject to a directional change in intensity or color. The first is used to learn the geometric properties of HEp-2 cells, while the second is used to learn the local intensity changes in HEp-2 cells. They were able to reach 98.27\% accuracy with the SNPHEp-2 \cite{ref_article10} dataset and 98.89 with the I3A dataset~\cite{ref_article10}.

2) CL-HEp2IC methods that use DNN as a classifier: \\
The following methods combine the feature extraction step and the classification step as one part. In fact, in this kind of method, the features of the input images are automatically extracted and relying on these features, the classification phase is achieved.\newline
The cell-level HEp2 image classification that uses DNN as a classifier is divided into three groups. The following methods combine the feature extraction step and the classification step as one part. In fact, in this kind of method, the features of the input images are automatically extracted and relying on these features, the classification phase is achieved. The cell-level HEp2 image classification that uses DNN as a classifier is divided into three groups. Generic DNN-based approaches exploit popular DNN architectures, primarily designed for general image classification tasks like ImageNet classification, to perform cell-level HEp2 image classification. The focus on this type of method began in 2015 with \cite{ref_article14} who used an AlexNet, a variant of CNN used in the competition of image classification of the ImageNet Database in 2012, reinforced by some pre-processing techniques such as image enhancement and data augmentation. They were able to achieve 80.3\% \cite{ref_article14} accuracy with the ICPR 2012 dataset. A later study conducted by \cite{ref_article15} concerning other generic CNN models, specially ImageNet, GoogleNet, and LetNet-5, shows that even without any pre-processing methods used in the procedure, GoogleNet exceeds the performance of the two remaining methods and gets as far as 95.53\% accuracy. In \cite{ref_article15}, the authors focused on three other generic CNNs, which are VGG-16, ResNet-50, and InceptionV3. Their work confirms that this last generic CNN, named InceptionV3 is more efficient in the HEp2 image classification than the other models considered in their research without any pre-processing procedures with a 98.28\% accuracy \cite{ref_article16}. As mentioned earlier, the generic CNN is strengthened with techniques such as image enhancing and DA, another idea brought by \cite{ref_article17} who suggested to use of the generative adversarial network (GAN), which generates images with a different composition than the initial images different from rotating or reversing, and train generic CNN like the GoogleNet model with those images provided. It reaches 98.6\% accuracy \cite{ref_article17}. Despite its high accuracy, this method shows its weaknesses while facing a large intra-class variation of data. Second, Generic DNN-based methods with partial changes in layers or training schemes consist of making smaller changes in the generic model. Two major studies were shown in this context: The first one is proposed by \cite{ref_article18}, who added two additional convolutional layers with a 1x1 filter before the LeNet5 CNN architecture. It aims to boost the number of feature channels to facilitate the task of the classification layers to attain a 79.13\% \cite{ref_article18} accuracy, which outperforms the accuracy of the basic LeNet-5 by a significant amount. The second method by the same group of researchers two years later in 2018: they chose to concentrate on the ResNet model and merge the first layers’ predictions into the final classification layers using a bridging mechanism to go far as 97.14\% accuracy on the ICPR 2012 dataset and 98.42\% \cite{ref_article20} accuracy on the I3A dataset. Despite the high accuracy achieved in this study, it is accompanied by an increasing number of network parameters.\\

\subsubsection{Specimen-level HEp-2 image classification (SL-HEP2IC) methods : }\leavevmode 
The specimen-level methods deal with the images of the HEP2 as an entire block and classify the whole image. These methods were decomposed into two main types: single-cell processing-based SL-HEP2IC methods and multi-cell processing-based methods.\\

1) Single-cell processing-based SL-HEP2IC methods: \\
Single-cell processing methods consist of feeding this kind of model with a specimen image as input. It will be split into singular cell images using ground truth labeling at the cell level, such as bounding box annotations or segmentation masks. Then a majority voting strategy is employed to obtain the specimen-level result by accumulating the cell results. It is considered the extension of the Cell-level HEp2 image classification.
According to the utilization of DNNs, we can split this part of the related work concerning the HEp2 image classification into two main parts:

\begin{itemize}
    \item Feature extraction-based methods: Those methods are much the same as those previously discussed. They are divided into two parts: feature extraction from each cell of the specimen and then passed into a classifier to obtain the cell class. In this context, the authors \cite{ref_article18} proposed a modification to the LeNet-5 model by adding a 1x1 convolutional block. Then, after the cell classes are identified, a majority voting strategy (MVS) is applied by choosing the most dominant cell label as the specimen label. This strategy leads them to reach 95.83\% accuracy \cite{ref_article18}. Confusion can be present while applying the majority voting strategy. That is why another approach is used to overcome such a critical issue: representing the population histogram (PH).
     \item Pure DNN-based methods: The pure DNN means that the DNN is used as a direct classifier in this kind of method. They are similar to the methods discussed previously.
    One of the studies concerning this type of method is the \cite{ref_article18} study, which used modified LeNet-5 for the classification issue for each cell image obtained from specimen images and reached a 79.13\% accuracy \cite{ref_article18}.
    \end{itemize}
The drawbacks of the single-cell processing based on specimen level are:
\begin{itemize}
        \item The requirement of single cell classification to obtain the specimen classification. 
        \item When numerous cells are presented in a specimen image, the single cell processing based on specimen level faces many problems while first of all segmenting and also while classifying specimen-cells.
    \end{itemize}

\begin{figure}[ht]
\centering
\frame{\includegraphics[width=5in ]{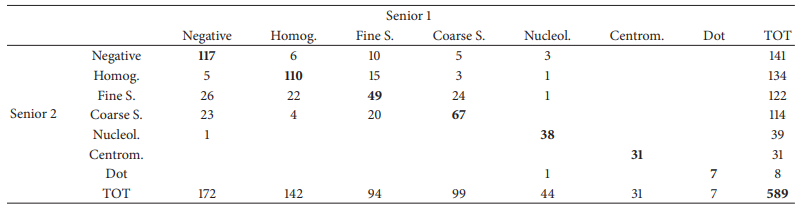}}
        \caption{Level of concordance between two seniors \cite{ref_article8}.}
\end{figure}

2) Multi-cell processing-based methods: \\
Multi-cell processing methods treat a whole specimen image at once. They are more efficient and can be divided into two types:
\begin{itemize}
    \item  Pixel-wise prediction-based methods: 
    In this prediction, each pixel of the specimen image is contributed to a class label. Then, the results of these pixels are fed to a majority voting method to classify the whole specimen. The basis of this method is the fully convolutional network, a variant of the CNN that replaces the fully connected layers with convolutional ones. 
    \item Image-wise prediction-based methods: In this type of method, only one prediction is given to the whole specimen image. A recent method is introduced in this context which resizes the specimen image. Consequently, the local information disappears. That is why handcrafted features are used to make the model stronger, showing a good classification performance that outperforms some other methods.

\end{itemize}
\section{Proposed approach}
In this section, the proposed approach is described, and we mention details about the dataset as well as model architectures.

\subsection{Immunofluorescence slides interpretation}
Medical decisions must be made carefully and with high precision. In this regard, the detection of the IIF image class is very crucial because of two factors :
\begin{itemize}
    \item The need for a double reading of these images, which is not always possible in many cases. 
    \item  Even with the availability of two senior physicians, the difficulty of analyzing IIF images is always present due to the great detail in this type of image. 
    \end{itemize}
Figure 3 visualizes the level of concordance between two seniors Immunologist readers. We note that they had, on average, 71\% of concordance despite their expertise in the field.

\subsection{Global approach process}
As shown in Figure 4, the user (who is either the doctor or the technician) must be able to:
\begin{itemize}
    \item  Identify whether the input image provided is positive or negative in the first instance.
    \item Provide prediction of the exact class of the IIF image if it is positive: The developed application, whose purpose is assisting Immunologists in diagnosing different cases, must be able to provide predictions with nearly perfect accuracy of the anomalies present in the patient's IIF images.
\end{itemize}

\begin{figure}
    \centering
    \includegraphics[width=1\textwidth]{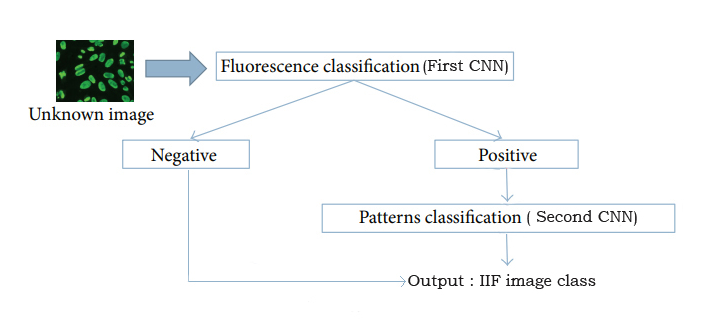}
    \caption{Proposed approach.}
    \label{fig1}
\end{figure}  

\subsection{Dataset}
For Binary classification, the dataset is a mixture of an open-source dataset called AIDA HEp-2 and a set of IIF images taken from the department of immunology of the military hospital of Tunis. The AIDA HEp-2 dataset is a part of the complete AIDA database, resulting from international cooperation between Tunisia and Italy.
In Table 1, the number of patients from whom the IIF images were taken is mentioned, as well as the positive and the negative patients. The total number of images acquired is equal to 2080, about 2 images from each patient.

\begin{table}
	\caption{Dataset Description}
	\centering
	\begin{tabular}{lll}
		\toprule                  \\
		Number of patients &  Positive fluorescence intensity & Negative fluorescence intensity  \\
		\midrule
		1000 &  470  & 530     \\
		
		\bottomrule
	\end{tabular}
	\label{tab:table}
\end{table}

For multi-class classification, a dataset containing 2668 IIF images distributed into 7 classes is used to train, validate and test the model. Figure 5 describes the representation of how these images are divided.

\begin{figure}
\centering
\includegraphics[width=3.9in ]{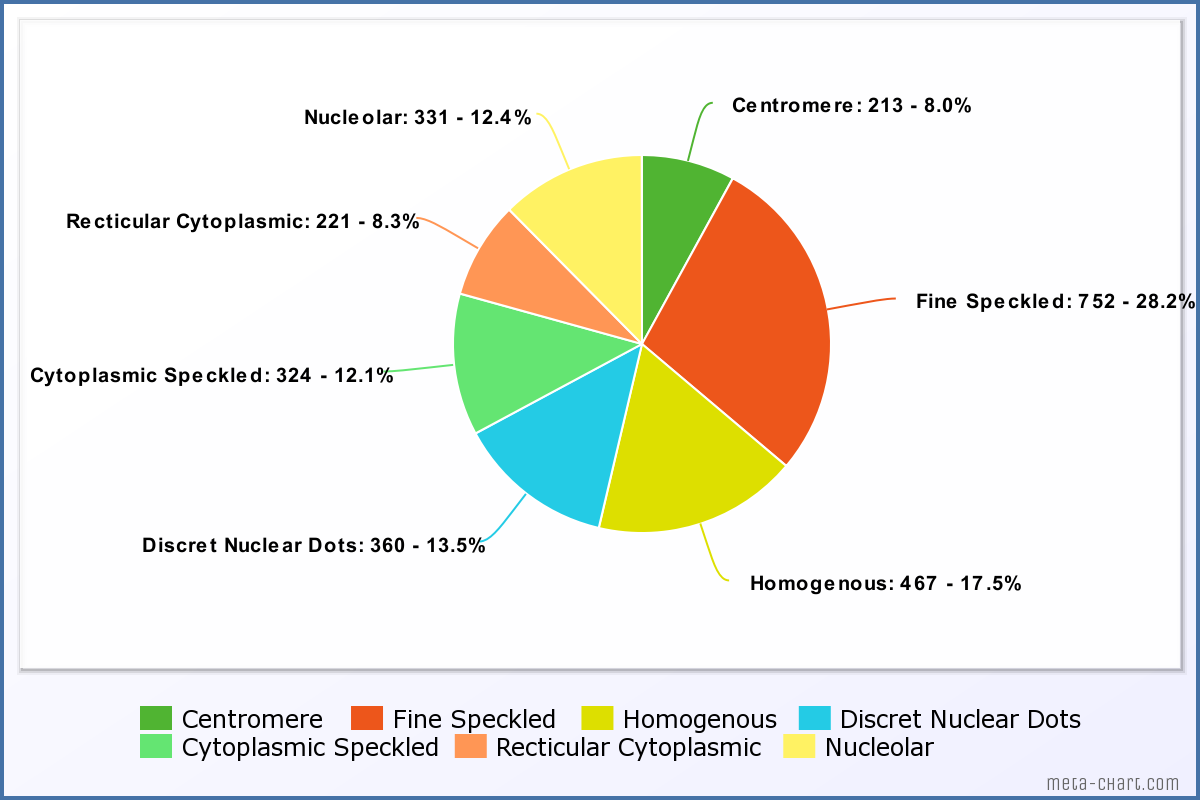}
\caption{Distribution of samples over classes.}
\end{figure}

\subsection{Model Architecture}
Two models were proposed for the prediction of the IIF image class. The first one is dedicated to the binary classification of the image provided as input: an IIF image is either positive (presence of antibody) or negative (otherwise). The model consists of multiple layers, one enveloping the other so that the output of the first becomes the input of the second and so on, until the last layer, which is the output of the whole model. In this work, the first model is sequential, i.e., the output of each layer is the input of the next layer. It is composed of the layers put in the following arrangement:
    \begin{itemize}
    \item 6 sets of Convolutional followed by a Max Pooling layer.
    \item A Flatten layer (with the shape of 256).
    \item 3 Dense layers (with the respective shapes of 64, 128, and 1).
    \end{itemize}

 For the second model, the proposed pre-trained NASNet neural network provides accurate results. We enhanced the architecture by adding a dense layer to the model which led to better results.

In this section, the obtained results through the training and testing phases are presented.

 \subsection{Training}
For the binary classification, a model was trained using two classes: positive and negative. Figure 6 shows a screen capture of results during the training phase. Then, the performance was evaluated  using a range of metrics. Next, the proposed approach was extended to the multiclass model, which was trained on a total of 7 different classes, as shown in Figure 7, and again evaluates the model's performance using a range of metrics.
 
 \subsubsection{Binary Model}
To investigate the performance of the proposed approach, we began by training a binary classification model with two classes: positive and negative. The positive class represented instances of the existence antibody in the cell, while the negative class represented instances that did not exhibit the phenomenon. A dataset of 2080 samples for training and testing the model and evaluating its performance was used.

    \begin{figure}
    \centering
    \includegraphics[width=4.5in]{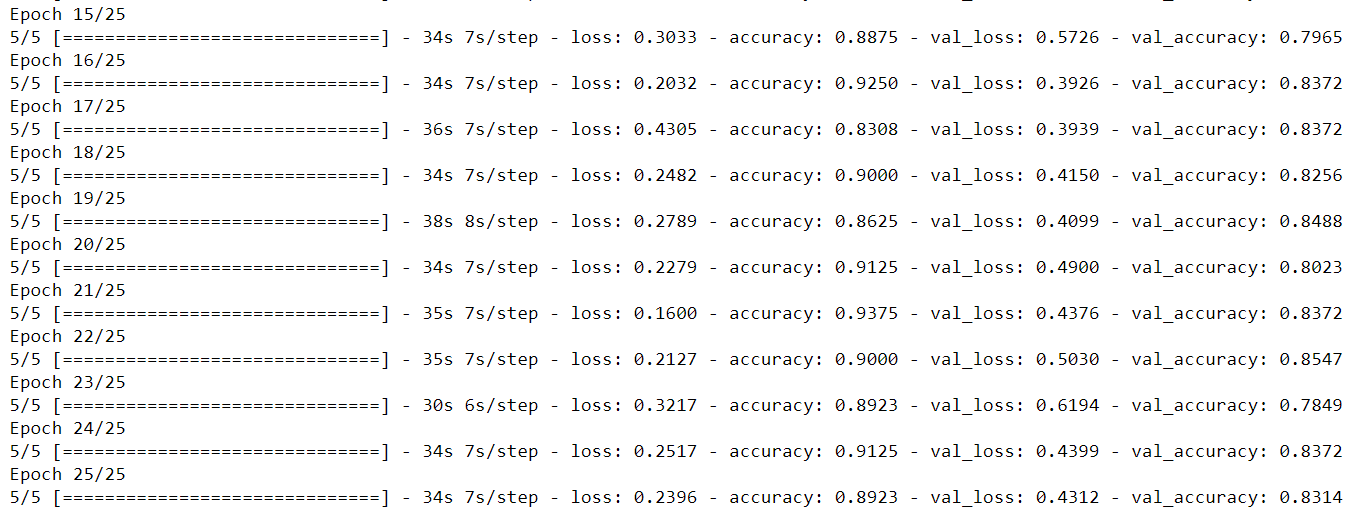}
    \caption{Training the binary classification model}
    \label{fig}
    \end{figure}
    
\subsubsection{Multi Class Classification}

A Multi-class model was trained to distinguish the different interpretable classes. There are seven classes, as explained previously, and the model was trained using Google Colab Pro as its training exceeded the locally provided resources.
    \begin{figure}
    \centering
    \includegraphics[width=4.5in]{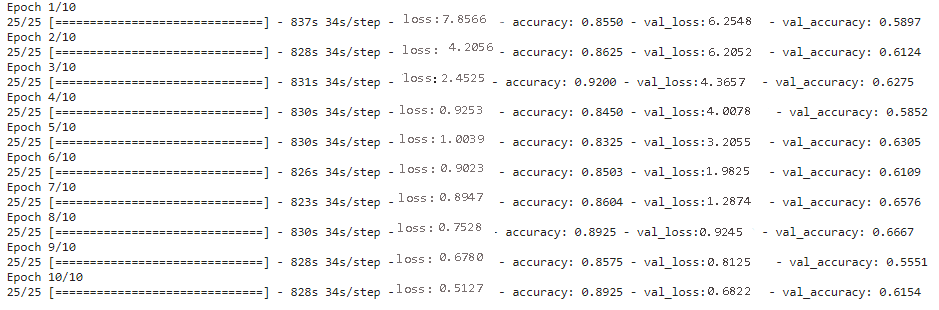}
    \caption{Training the multi-class classification model}
    \label{fig}
    \end{figure}
\subsection{Testing Results}

The first model achieved an accuracy of 94.48\% on a challenging classification task. The model was trained using a convolutional neural network architecture as explained in the previous section on a dataset of 2080 images. Then it was evaluated using a standard test set and obtained a precision-recall curve, which showed that the model's accuracy was consistent across a range of thresholds. The results of this study demonstrate the effectiveness of this approach and provide a foundation for future research.

The second model achieved an accuracy of 69.61\% on the multiclass classification task. The second model was trained using a more complex neural network architecture with additional layers and a larger number of parameters. A similar dataset of 2668 images was used for training and testing the second model. The evaluation of the second model's performance showed that it had lower accuracy compared to the first model, which may be due to the lack of images or the increased complexity of the architecture.
\section{Conclusion}

In this work, an overview of the different state-of-the-art techniques, methods, and  tools related to AI for the Interpretation of Immunofluorescence slides is depicted. Also, a solution that treats the multiple cases of immunofluorescence slide interpretation is presented using a combination of a DL model and a NASNet model leading to accurate results, especially in binary classification. In future works, a fusion between the two methods, cell level and specimen level, is a must-try approach to reach better results. In fact, with the use of the cell-level method, a population diagram representing the distribution of the cells within the sample can be produced. Then, with the use of the prediction based on the specimen-level method, a comparison of the two results will lead to a better verification of the decision-making process. Finally, we note that AI is a powerful tool in the Immunology field, however, human beings are exposed to a variety of diseases that target their immunity, therefore, new AI solutions evolving with theses diseases are required.

\section*{Acknowledgment}
The authors would like to thank Prince Sultan University for financially supporting the conference attendance fees.


\begin{thebibliography}{8}
\bibitem{ref_article89}
Bagatini Al., 2018. Immune system and chronic diseases 2018. Journal of immunology research, 2018.

\bibitem{ref_article90}
Van Hoovels Al. "Variation in antinuclear antibody detection by automated indirect immunofluorescence analysis." Annals of the Rheumatic Diseases 78, no. 6 (2019): e48-e48.

\bibitem{ref_article91}
Cascio Al. "Deep CNN for IIF images classification in autoimmune diagnostics." Applied Sciences 9, no. 8 (2019): 1618.

\bibitem{ref_article92}
Jain Al. "Role of direct immunofluorescence microscopy in spectrum of diffuse proliferative glomerulonephritis: A single-center study." Journal of Microscopy and Ultrastructure 9, no. 4 (2021): 177.


\bibitem{ref_article2}
Olivier Boyer and Sophie Candon, F.:Autoimmune diseases: The breakdown of self-tolerance. \url{https://www.inserm.fr/information-en-sante/dossiers-information/maladies-auto-immunes} \textbf (2021)

\bibitem{ref_article3}
Linlin Shen, Jiaming Lin Al.: Hep-2 image
classification using intensity order pooling based features and bag of
words. Pattern Recognition, 47(7):2419–2427, (2014)

\bibitem{ref_article4}
Lingqiao Liu and Lei Wang F.: Hep-2 cell image classification with mul-
tiple linear descriptors. Pattern Recognition,47(7):2400–2408\textbf{2} (2014)
\bibitem{qawqzeh2020classification}Qawqzeh, Y., Bajahzar, A., Jemmali, M., Otoom, M. \& Thaljaoui, A. Classification of diabetes using photoplethysmogram (PPG) waveform analysis: Logistic regression modeling. {\em BioMed Research International}. \textbf{2020} (2020)

\bibitem{driss2020novel}Driss, K., Boulila, W., Batool, A. \& Ahmad, J. A novel approach for classifying diabetes’ patients based on imputation and machine learning. {\em 2020 International Conference On UK-China Emerging Technologies (UCET)}. pp. 1-4 (2020)
\bibitem{al2021novel}Al-Sarem, M., Alsaeedi, A., Saeed, F., Boulila, W. \& AmeerBakhsh, O. A novel hybrid deep learning model for detecting COVID-19-related rumors on social media based on LSTM and concatenated parallel CNNs. {\em Applied Sciences}. \textbf{11}, 7940 (2021)
\bibitem{al2021feature}Al-Sarem, M., Saeed, F., Boulila, W., Emara, A., Al-Mohaimeed, M. \& Errais, M. Feature selection and classification using CatBoost method for improving the performance of predicting Parkinson’s disease. {\em Advances On Smart And Soft Computing: Proceedings Of ICACIn 2020}. pp. 189-199 (2021)
\bibitem{ben2022randomly}Ben Atitallah, S., Driss, M., Boulila, W. \& Ben Ghezala, H. Randomly initialized convolutional neural network for the recognition of COVID-19 using X-ray images. {\em International Journal Of Imaging Systems And Technology}. \textbf{32}, 55-73 (2022)
\bibitem{rasool2022hybrid}Rasool, M., Ismail, N., Boulila, W., Ammar, A., Samma, H., Yafooz, W. \& Emara, A. A Hybrid Deep Learning Model for Brain Tumour Classification. {\em Entropy}. \textbf{24}, 799 (2022)

\bibitem{jemmali2022equity}Jemmali, M., Melhim, L., Alourani, A. \& Alam, M. Equity distribution of quality evaluation reports to doctors in health care organizations. {\em PeerJ Computer Science}. \textbf{8} pp. e819 (2022)

\bibitem{alam2022public}Alam, M., Melhim, L., Ahmad, M. \& Jemmali, M. Public attitude towards covid-19 vaccination: validation of covid-vaccination attitude scale (c-vas). {\em Journal Of Multidisciplinary Healthcare}. pp. 941-954 (2022)

\bibitem{ref_dirindir}
https://www.abcam.com/secondary-antibodies/direct-vs-indirect-immunofluorescence


\bibitem{ref_article5}
Xian-Hua Han Al. Hep-2 cell classification using k-support spatial pooling in deep cnns. In Deep Learning and Data Labeling for Medical Applications, pages 3–11. Springer, 2016

\bibitem{ref_article6}
Ha Tran Hong Phan Al: Transfer learning of a convolutional neural network for hep-2 cell
image classification. In 2016 IEEE 13th International Symposium on
Biomedical Imaging (ISBI), pages 1208–1211. IEEE, (2016)
\bibitem{ref_article7}
Mengchi Lu Al. Hep-
2 cell image classification method based on very deep convolutional
networks with small datasets. In Ninth International Conference on Dig-
ital Image Processing (ICDIP 2017), volume 10420, page 1042040. Inter-
national Society for Optics and Photonics, (2017)
\bibitem{ref_article8}
Amel Benammar Elgaaied Al. Computer-assisted classification patterns in autoimmune di-
agnostics: the aida project. BioMed research international, (2016).
\bibitem{ref_article10}
Caleb Vununu, Suk-Hwan Lee, Oh-Jun Kwon, and Ki-Ryong Kwon. A dynamic learning method for the classification of the hep-2 cell images. Electronics, 8(8):850, 2019
\bibitem{ref_article14}
Neslihan Bayramoglu Al. Human epithelial type 2 cell classification with convolutional neural networks. In 2015 IEEE 15th International Conference on Bioinformatics and Bioengi-
neering (BIBE), pages 1–6. IEEE, 2015
\bibitem{ref_article15}
Larissa Ferreira Rodrigues Al. Hep-2 cell image classification based on convolutional neural networks. In 2017 Workshop of Computer Vision (WVC), pages 13–18. IEEE,2017
\bibitem{ref_article16}
Larissa Ferreira Rodrigues Al. Comparing convolutional neural networks and preprocessing techniques for hep-2 cell classification in immunofluorescence images. Computers in biology and medicine, 116:103542, 2020
\bibitem{ref_article17}
Tomáš Majtner Al. On the effectiveness of generative
adversarial networks as hep-2 image augmentation tool. In Scandinavian Conference on Image Analysis, pages 439–451. Springer, 2019.
\bibitem{ref_article18}
Hongwei Li Al. Deep cnns for hep-2 cells classification: A cross-specimen analysis. arXiv preprint
arXiv:1604.05816, 2016.
\bibitem{ref_article19}
Siqi Li, Huiyan Jiang, and Wenbo Pang. Joint multiple fully connected convolutional neural network with extreme learning machine for hepatocellular carcinoma nuclei grading. Computers in biology and medicine, 84:156–167, 2017
\bibitem{ref_article20}
Haijun Lei, Tao Han, Feng Zhou, Zhen Yu, Jing Qin, Ahmed Elazab, and Baiying Lei. A deeply supervised residual network for hep-2 cell
classification via cross-modal transfer learning. Pattern Recognition, 79:290–302, 2018
\bibitem{ref_article21}
Jingxin Liu, Bolei Xu, Linlin Shen, Jon Garibaldi, and Guoping Qiu. Hep-2 cell classification based on a deep autoencoding-classification convolutional neural network. In 2017 IEEE 14th International Symposium on Biomedical Imaging (ISBI 2017), pages 1019–1023. IEEE, 2017
\bibitem{ref_article22}
Yuexiang Li and Linlin Shen. A deep residual inception network for hep-2 cell classification. In Deep Learning in Medical Image Analysis and Multimodal Learning for Clinical Decision Support, pages 12–20. Springer, 2017.
\end{thebibliography}
\end{document}